\documentclass{article}
\usepackage{graphicx} 
\usepackage[backend=biber,style=apa]{biblatex}
\usepackage{float}

\title{Shifts in U.S. Social Media Use, 2020–2024: Decline, Fragmentation, and Enduring Polarization}
\author{Petter Törnberg}
\date{October 2025}

\usepackage{amsmath}
\usepackage[font=small,labelfont=bf]{caption}
\usepackage[font=small,labelfont=bf]{subcaption}

\begin{document}
\maketitle

\begin{abstract}
\noindent Using nationally representative data from the 2020 and 2024 \textit{American National Election Studies} (ANES), this paper traces how the U.S.\ social media landscape has shifted across platforms, demographics, and politics. Overall platform use has declined, with the youngest and oldest Americans increasingly abstaining from social media altogether. Facebook, YouTube, and Twitter/X have lost ground, while TikTok and Reddit have grown modestly, reflecting a more fragmented digital public sphere. Platform audiences have aged and become slightly more educated and diverse. Politically, most platforms have moved toward Republican users while remaining, on balance, Democratic-leaning. Twitter/X has experienced the sharpest shift: posting has flipped nearly 50 percentage points from Democrats to Republicans. Across platforms, political posting remains tightly linked to affective polarization, as the most partisan users are also the most active. As casual users disengage and polarized partisans remain vocal, the online public sphere grows smaller, sharper, and more ideologically extreme.
\end{abstract}

\section{Introduction}
The social media ecosystem appears to be in flux. Twitter’s rebranding to X has come to symbolize a broader reorganization of online publics -- political reshuffling, user attrition, and uncertainty over the role of legacy platforms. Simultaneously, the shift from text-based, networked feeds toward algorithmically curated, short-form video has accelerated, with TikTok setting the pace for the wider ecosystem \parencite{zulli2022extending,gillespie2014relevance,kaye2021co}. At the same time, everyday communication increasingly migrates from large, open networks to semi-private spaces such as group chats and messaging apps. Commentators describe a digital public sphere in transition: smaller, more fragmented, less dominated by traditional social networking sites, and rise of more broadcast-oriented forms of media -- with some even suggesting that we are seeing the beginning of the end of the social media era. 

Despite this abundant commentary, we know surprisingly little about these changes from representative data. Most accounts rely on commercial analytics or opt-in web surveys that capture broad signals but not population-level patterns \parencite{blank2017representativeness,hargittai2020potential}. The \textit{American National Election Studies} (ANES) now provides rare, comparable measures of platform use in 2020 and 2024 \parencite{anes2020,anes2024}, including detailed items on frequency and political posting for Facebook and Twitter/X. These data make it possible to trace, with national representativeness, how social media use in the United States has shifted across platforms, demographics, and political orientations over a politically turbulent period.

This paper provides a descriptive overview of these shifts. It first documents the modest but clear decline in overall platform reach and the growing share of Americans who report using no social media at all—especially among the youngest and oldest cohorts. It then examines how platform audiences have aged and become somewhat more educated and racially diverse. Finally, it turns to the political dimension, showing that Twitter/X has moved from a Democratic-leaning to a Republican-leaning user base, particularly among frequent posters, and that political posting across Facebook and Twitter/X remains strongly linked to affective polarization.

Taken together, these patterns portray a social media environment in gradual contraction and segmentation. As participation becomes more uneven and politically polarized users remain the most active contributors, visible online discourse continues to amplify partisan extremes—even as the broader public quietly disengages.

\section{Methods}

\subsection{Data}
The analyses draw on the 2020 and 2024 \textit{American National Election Studies} (ANES) Time Series surveys. Both waves are nationally representative of U.S.\ citizens aged 18+, based on address-based probability samples administered online and face-to-face in English and Spanish. The 2020 survey included 8{,}280 pre-election and 7{,}449 post-election respondents; the 2024 survey included 5{,}521 pre-election and 4{,}964 post-election reinterviews.

The ANES sampling frame covers U.S.\ citizens residing in households across all 50 states and the District of Columbia. By design, it excludes non-citizens, institutionalized populations (e.g., prisons or long-term care facilities), active-duty military abroad, and individuals without fixed addresses. In practice, non-English/non-Spanish speakers and those lacking reliable internet access are also underrepresented. Findings therefore describe social media use among U.S.\ citizens rather than all adults living in the United States.

To ensure comparability across years, all analyses employ the post-election full-sample weights (\texttt{V200010b} for 2020; \texttt{V240107b} for 2024). These weights adjust for unequal selection probabilities and nonresponse via iterative raking on age~$\times$~gender and race~$\times$~education margins. Weighted estimates approximate population distributions of U.S.\ citizens; all figures report weighted means or proportions. 

\subsection{Variables}

\paragraph{Demographics.}
Respondent sex (\texttt{V201600}/\texttt{V241550}), age (\texttt{V201507x}/\texttt{V241458x}), race/ethnicity (\texttt{V201549x}/\texttt{V241501x}), and education (\texttt{V201511x}/\texttt{V241465x}) were harmonized into common categories: \textit{White non-Hispanic}, \textit{Black}, \textit{Hispanic}, \textit{Asian}, and \textit{Other/Mixed} for race; and \textit{High school or less}, \textit{Some college}, \textit{Bachelor’s}, and \textit{Postgraduate} for education.

\paragraph{Social media use.}
Platform visitation was measured with dichotomous indicators for whether respondents used Facebook, Twitter/X, Instagram, Reddit, YouTube, Snapchat, or TikTok (\texttt{V202541a–g} in 2020; \texttt{V242577a–g} in 2024). Respondents who answered all items and reported no use were coded as \textit{non-users}.

\paragraph{Political activity on social media.}
For Facebook and Twitter/X, the ANES asked about frequency of use and political posting (\texttt{V202542}/\texttt{V202545} and \texttt{V242578} / \texttt{V242581}). Responses were recoded to continuous scales from 0 (never) to 1 (most frequent) for comparability across years.

\paragraph{Users, visits, and posts.}
To capture how partisanship varies with activity, the analysis distinguishes three engagement tiers: (1) all users, (2) visit-weighted users, and (3) post-weighted users. Partisanship was measured via presidential vote choice, recoded into Democrat, Republican, or Other, and summarized as the net difference $(Dem - Rep)$ in percentage points.

For engagement weighting, survey weights were multiplied by frequency-derived intensity scores. Visit frequency categories (``multiple times per day'' to ``less than once a week'') were mapped to approximate monthly visit counts (35, 21, 7, 3, 1, 0.5, 0.2); posting frequencies (``multiple times per day'' to ``never'') to 20, 10, 5, 2, 0. These adjusted weights approximate the partisan balance of (a) platform audiences, (b) engagement, and (c) visible content, illustrating how composition shifts from broad users to the most active participants.

\paragraph{Affective polarization.}
Affective polarization was defined as the absolute difference between respondents’ feeling-thermometer ratings of Republicans and Democrats:
\[
\left|\mathrm{FT}_{\rm Rep} - \mathrm{FT}_{\rm Dem}\right|.
\]
For visualization, respondents were grouped into eight equal-width bins (0–100), and weighted mean use and posting frequencies were computed within each bin for 2020 and 2024. Confidence intervals were estimated via nonparametric bootstrap.

\paragraph{Political affiliation.}
Presidential vote choice was coded as Democrat, Republican, or Other. Party identification followed the standard three-category ANES coding (\texttt{pid3}).

\subsection{Analytical approach}
Analyses were conducted in Python using \texttt{pandas}, \texttt{numpy}, and \texttt{matplotlib}. Weighted point estimates and 95\% confidence intervals were computed using post-election sample weights, with sampling uncertainty adjusted via Kish’s effective sample size correction to account for unequal weighting.

To reduce sensitivity to extreme weights, robustness checks applied 1–2\% winsorization of the weight distribution. Results were substantively unchanged, though trimming produced slightly narrower confidence intervals for small subgroups (e.g., platform-specific user bases).

All analyses are descriptive and aim to trace compositional change rather than causal relationships. Harmonized coding across waves enables consistent longitudinal comparisons of the demographic and political composition of major social media platforms.

\section{Results}

\subsection{Declining Use of Social Media}

\begin{figure}[H]
\centering
\begin{subfigure}[t]{0.32\textwidth}
  \centering
  \includegraphics[width=\linewidth]{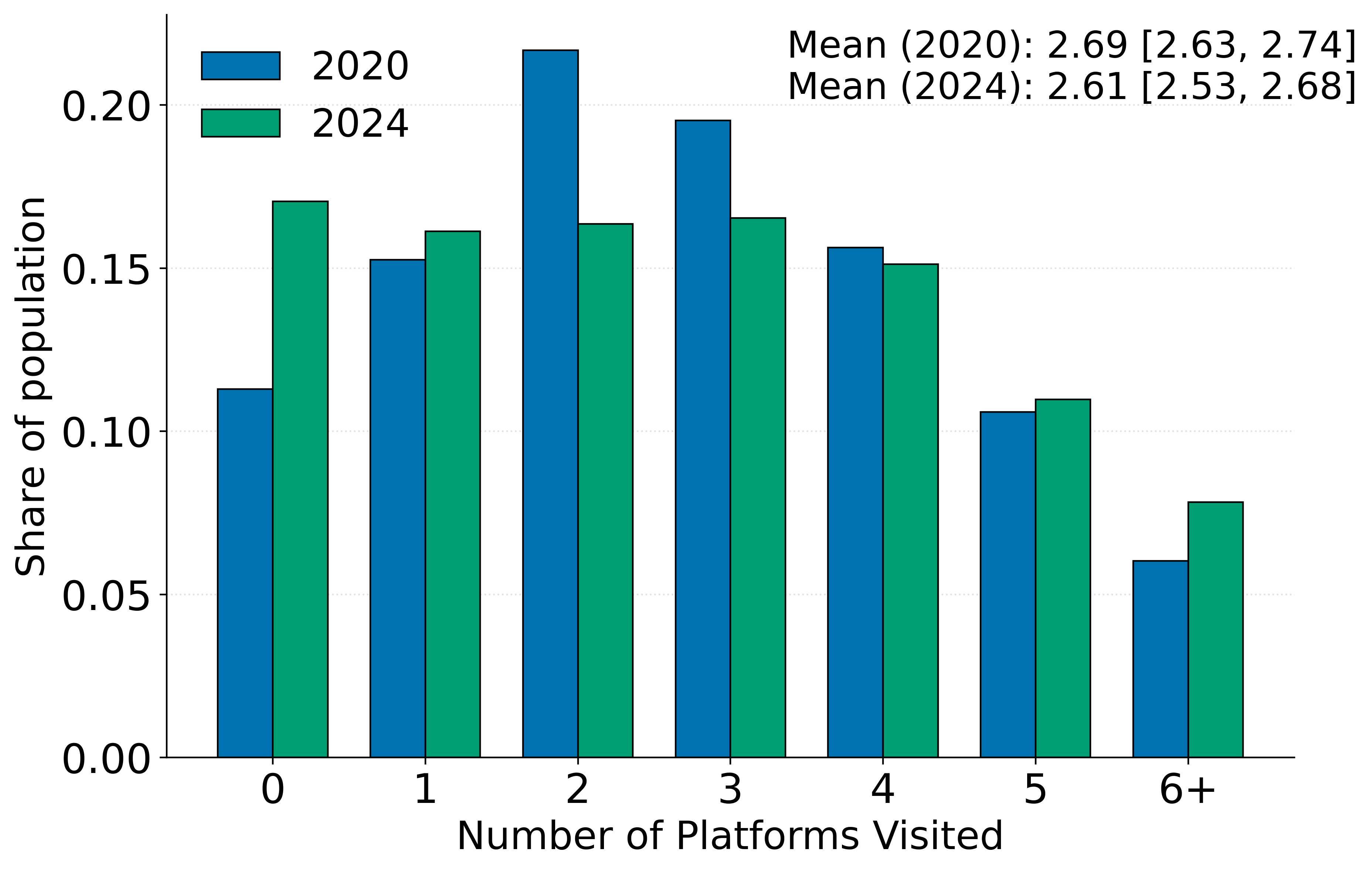}
  \caption{Number of platforms used (ANES 2020 vs.\ 2024, weighted).}
  \label{fig:num_platforms}
\end{subfigure}\hfill
\begin{subfigure}[t]{0.32\textwidth}
  \centering
  \includegraphics[width=\linewidth]{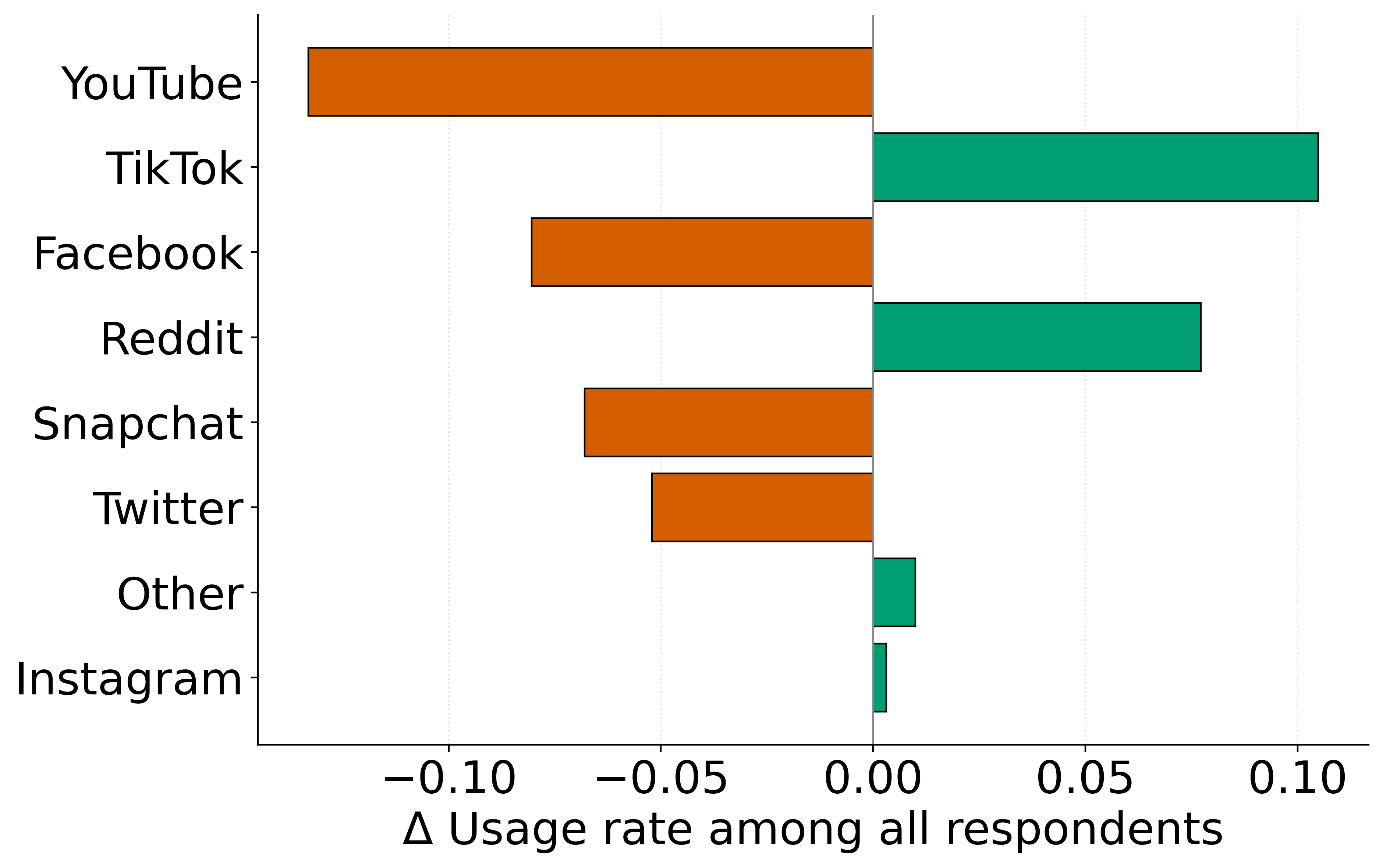}
  \caption{Net change in platform reach (2024--2020, share of all respondents).}
  \label{fig:delta_usage}
\end{subfigure}\hfill
\begin{subfigure}[t]{0.32\textwidth}
  \centering
  \includegraphics[width=\linewidth]{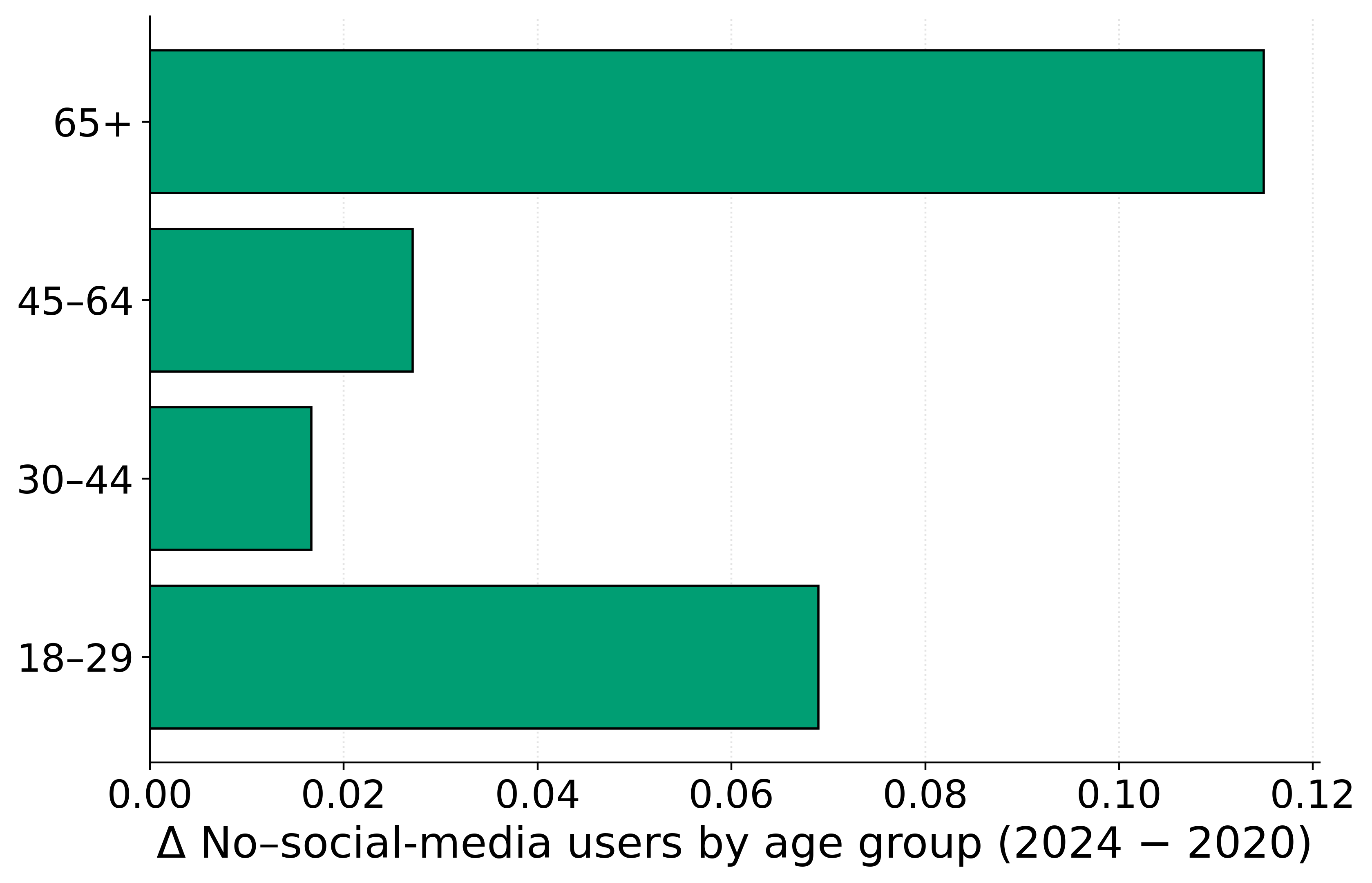}
  \caption{Change in non-use by age group (2024--2020, 95\% CI).}
  \label{fig:no_platform_age}
\end{subfigure}
\caption{Shifts in overall social media engagement, 2020--2024 (ANES, weighted population estimates with 95\% CIs). The share of respondents reporting no platform use increased clearly between 2020 and 2024, while most major platforms—YouTube, Facebook, Snapchat, and Twitter/X—lost reach. TikTok and Reddit expanded modestly, indicating consolidation around short-form video and discussion-based platforms.}
\label{fig:decline_overview}
\end{figure}

Between 2020 and 2024, the average number of social media platforms used declined slightly, from 2.69~[2.63,~2.74] to 2.61~[2.53,~2.68] (Fig.~\ref{fig:num_platforms}). This reduction is modest but notable, driven primarily by a growing share of respondents who report using no social media at all. At the same time, the right tail of heavy multi-platform users expanded slightly, suggesting that both non-use and high-intensity use have increased, while mid-level participation (two to three platforms) has contracted. The overall pattern points to a widening polarization in participation between heavy users and those opting out entirely.

The growth in non-use are especially pronounced at the age extremes (Fig.~\ref{fig:no_platform_age}). The share of adults aged 65~and older reporting no social media use increased by roughly 12 percentage points, while non-use among those aged 18–29 rose by around 7 points. Middle-aged groups (30–44 and 45–64) show marginal increases. This pattern underscores that both younger and older cohorts are withdrawing from social media. 

Platform-specific trends (Fig.~\ref{fig:delta_usage}) reinforce this picture. Traditional social network platforms and long-form video sites have declined: YouTube, Facebook, Snapchat, and Twitter/X all show reduced reach, while TikTok and Reddit record modest gains and Instagram remains stable. Despite the emergence of new entrants such as Threads and Bluesky, the ``Other'' category has not grown significantly. Taken together, these shifts indicate a gradual reorganization of the ecosystem, with short-form video and discussion-oriented platforms gaining ground as legacy networks lose momentum.

\subsection{Partisan Reconfiguration: Twitter/X Posting Flips Republican}

\begin{figure}[H]
\centering
\begin{subfigure}[t]{0.95\textwidth}
\includegraphics[width=0.9\textwidth]{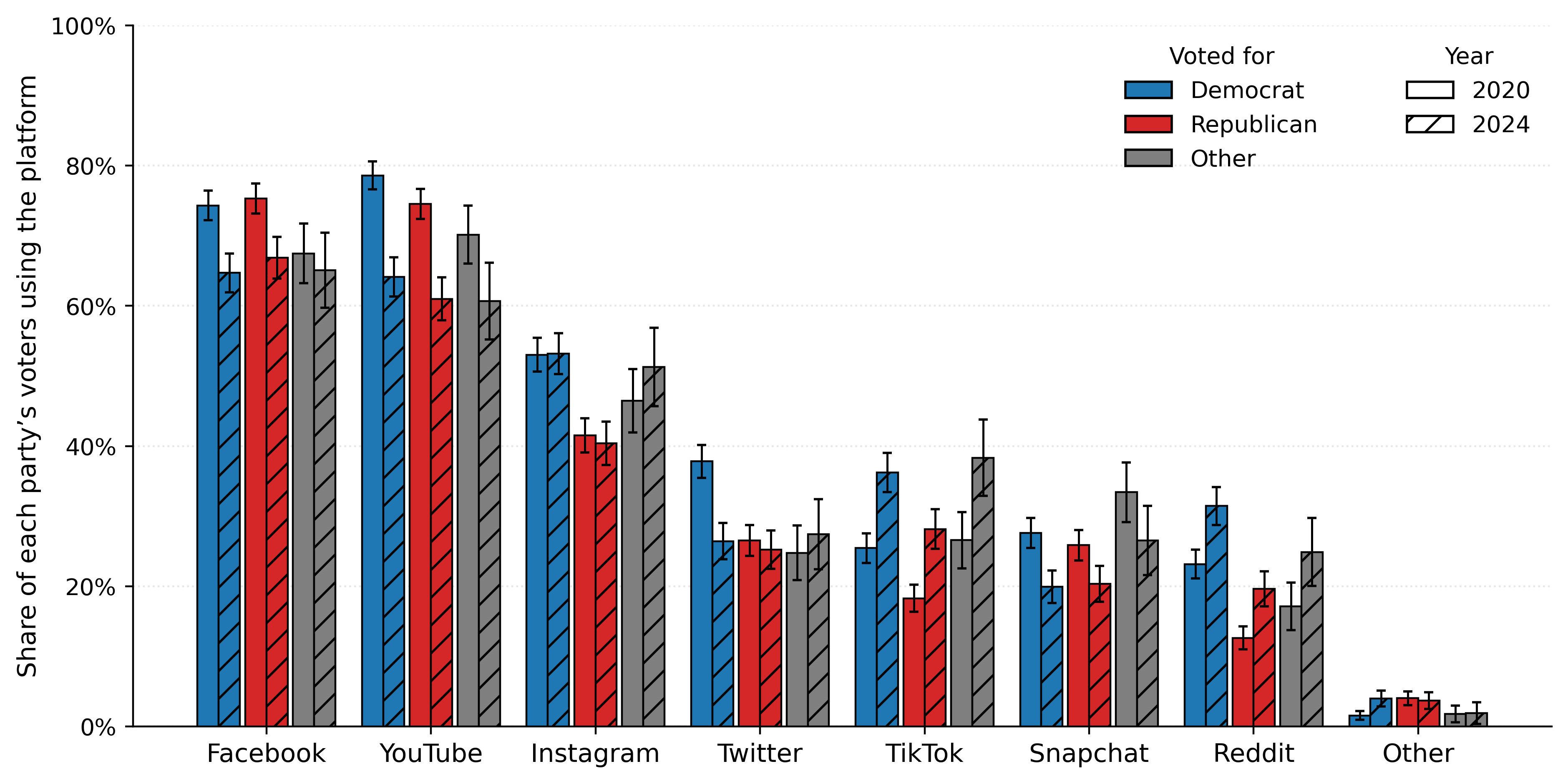}
\caption{Platform usage by presidential vote choice, 2020 vs.\ 2024 (ANES, weighted population estimates with 95\% CIs). Facebook and YouTube remain the largest platforms across all voter groups, despite declining reach. TikTok and Reddit expand modestly among both Democratic and Republican voters, whereas Twitter/X loses Democratic users but retains its Republican base.}
\label{fig:vote_usage}
\end{subfigure}\hfill
\centering
\begin{subfigure}[t]{0.6\textwidth}
  \centering
  \includegraphics[width=\linewidth]{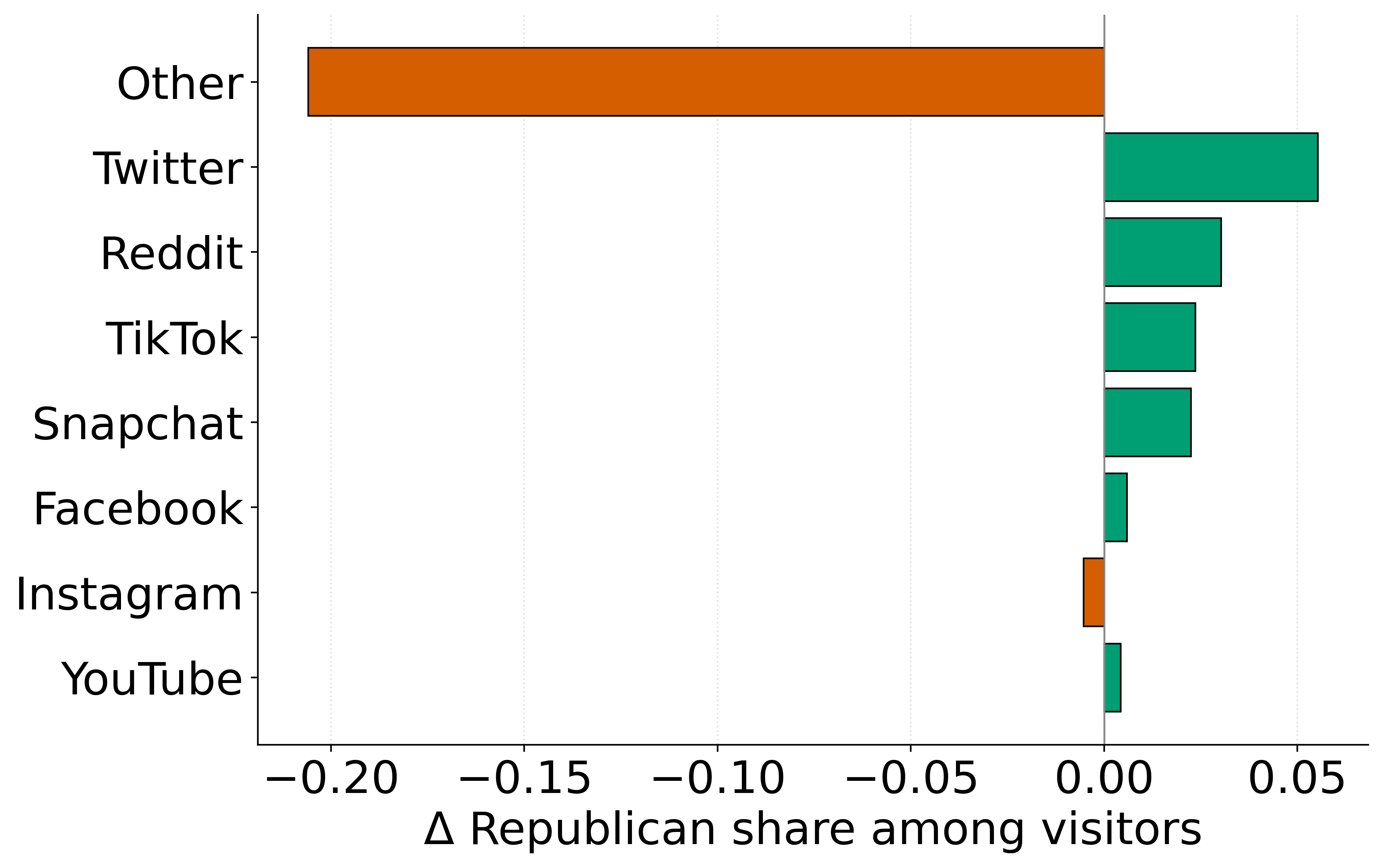}
  \caption{Change in Republican share among platform visitors, 2024–2020.}
  \label{fig:delta_rep}
\end{subfigure}\hfill
\begin{subfigure}[t]{0.35\textwidth}
  \centering
  \includegraphics[width=\linewidth]{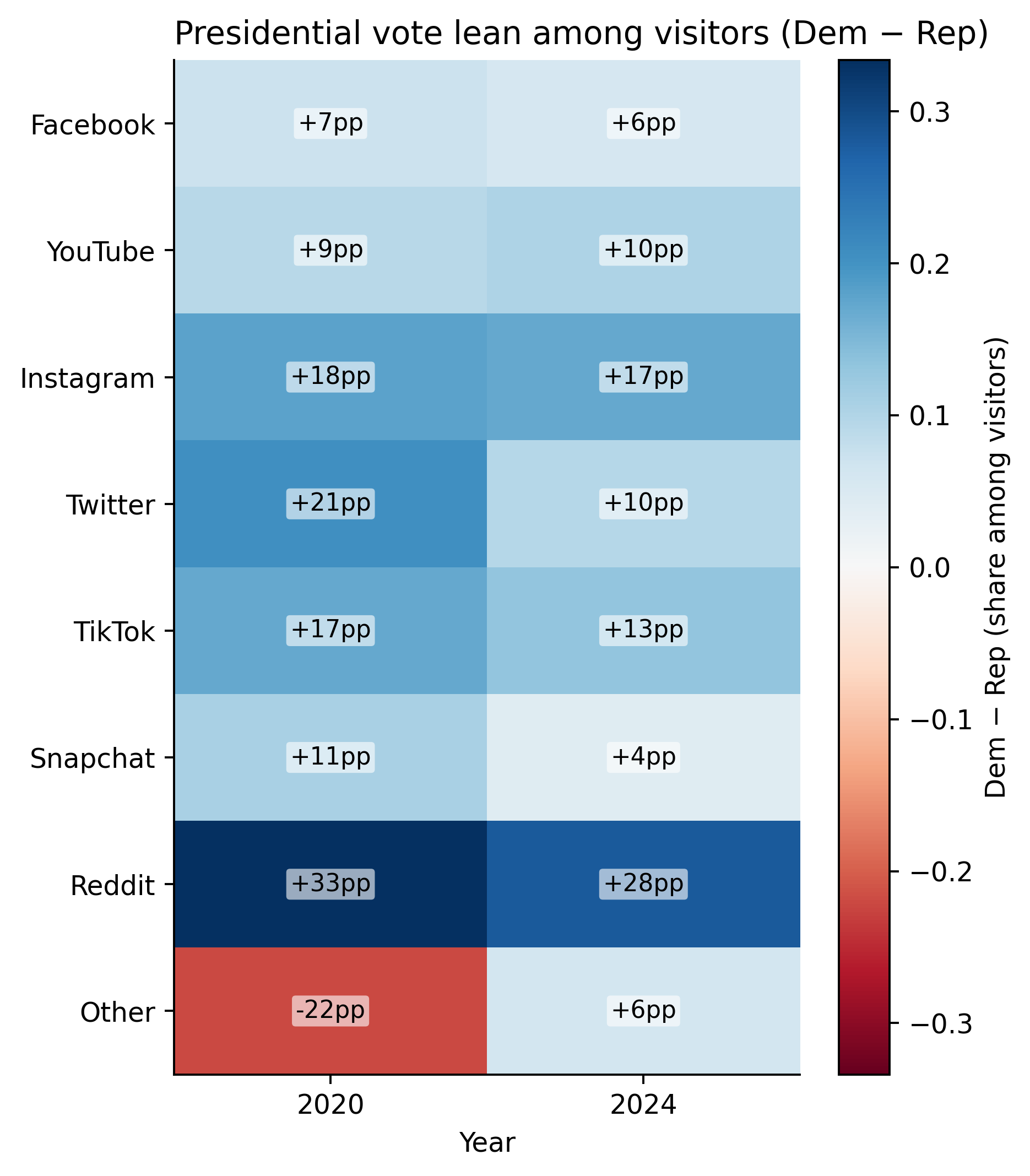}
  \caption{Democratic–Republican vote lean (percentage-point difference).}
  \label{fig:vote_lean}
\end{subfigure}
\caption{Directional shifts in partisan balance across major platforms, 2020–2024 (ANES, weighted). Twitter/X and Reddit show the strongest Republican gains, while Democratic overrepresentation persists—though more moderately—on TikTok and Reddit.}
\label{fig:partisan_shifts}
\end{figure}

Figure~\ref{fig:vote_usage} shows that, despite their declining reach, Facebook and YouTube remain the dominant platforms across voter groups. Twitter/X use has fallen sharply among Democratic voters and is now surpassed by TikTok. Smaller platforms continue to attract only relatively limited audiences.

Examining changes in the Republican share of users, Figure~\ref{fig:partisan_shifts} reveals that nearly all major platforms have shifted toward Republicans, though at varying rates. The Republican share rises most steeply on Twitter/X and, to a lesser extent, Reddit and TikTok. While Figure~\ref{fig:delta_usage} indicated that the ``Other'' category has not grown in its user base, Figure~\ref{fig:partisan_shifts} suggests substantial internal realignment, with a roughly 20~percentage-point move toward Democratic users.

Despite these shifts toward more Republican users, Figure~\ref{fig:vote_lean} shows that all major social media platforms remain, on balance, Democratic-leaning. Reddit and TikTok retain the largest Democratic advantages, while Twitter/X now approaches partisan parity. Overall, partisan asymmetries across platforms have narrowed: the formerly strong Democratic skew of the social media landscape has weakened, producing a more ideologically balanced environment in terms of user base.

While mainstream platforms have become relatively more Republican, the ``Other'' category has moved in the opposite direction, from strongly Republican-leaning to Democratic-leaning. This pattern points to a twin movement: Republican users shifting from ideologically homogeneous venues such as TruthSocial into mainstream platforms such as Twitter/X, whereas Democrats have retreated toward emerging, smaller networks such as Bluesky, Mastodon, and Threads. The result is a more fragmented digital ecosystem, with partisan groups redistributing across old and new spaces.

\begin{figure}[H]
    \centering
    \includegraphics[width=0.85\textwidth]{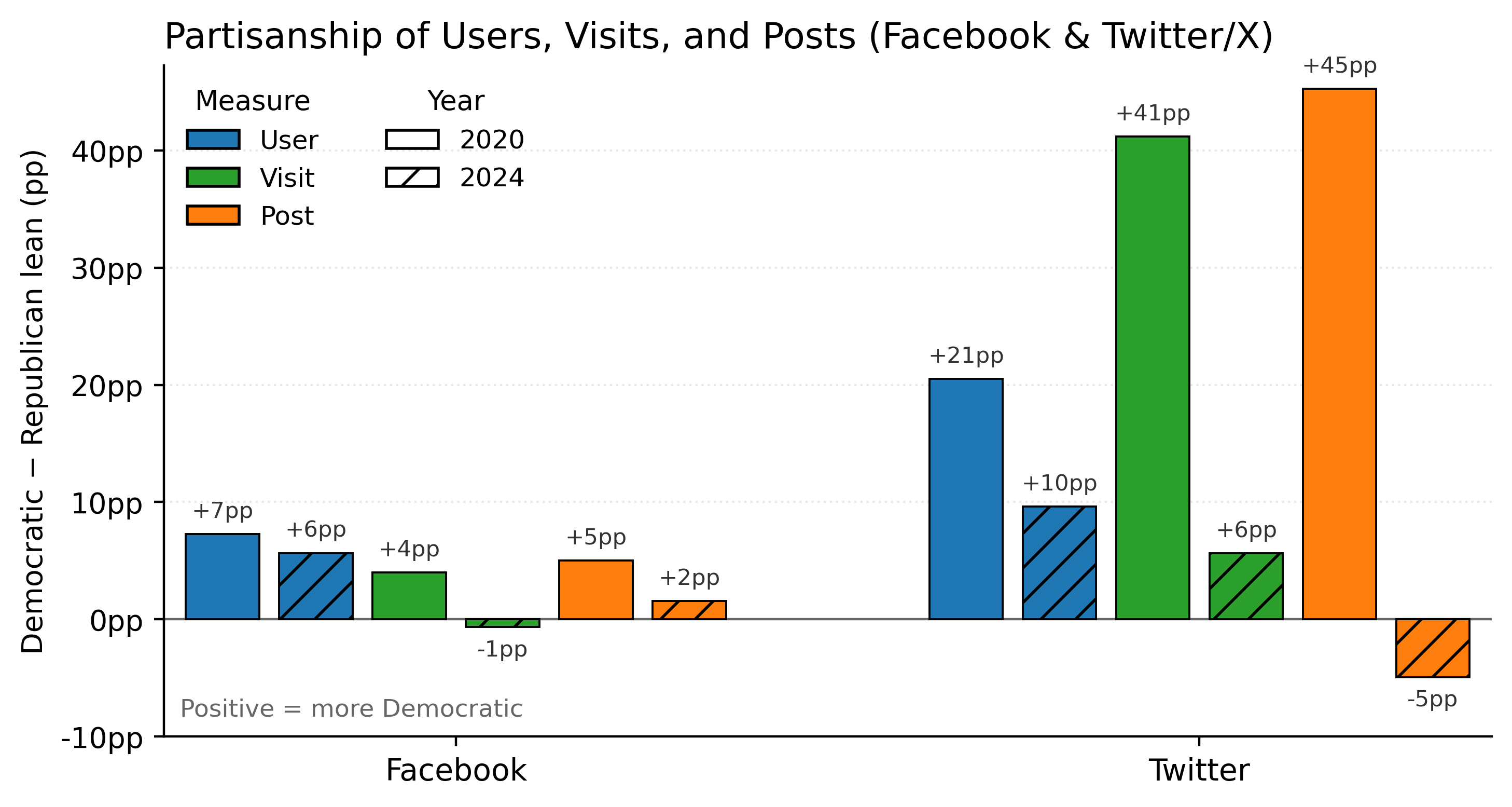}
    \caption{Shift in partisan composition of users, visits, and posts on Facebook and Twitter/X, 2020–2024 (ANES, weighted). Data on posting and use frequency are available for Twitter/X and Facebook only. Values show the difference between the shares of Democratic and Republican voters (Dem – Rep, percentage points). User-level estimates represent the average partisan balance among all platform users, while visit- and post-weighted measures emphasize the composition of more active participants.}
    \label{fig:partisanship_fb_twitter}
\end{figure}

The focus on users alone however conceals who actually dominates platform activity, which depends more on engagement frequency. Figure~\ref{fig:partisanship_fb_twitter} disaggregates the partisan balance of Facebook and Twitter/X (detailed data are only available for these platforms) by intensity -- comparing all users, frequent visitors, and frequent posters (see Methods). Facebook remains comparatively stable but exhibits a modest Republican shift among use and posting. 

Twitter/X, by contrast, undergoes a dramatic reversal. In 2020, Democrats outnumbered Republicans across all engagement levels and overwhelmingly dominated posting activity -- with a roughly 45~percentage-point Democratic advantage among posts. By 2024, this pattern has flipped: while the user base remains slightly Democratic-leaning, visit-weighted estimates show radically reduced Democratic presence, and post-weighted estimates reveal a complete inversion, with a modest Republican majority. This represents an approximately 50~percentage-point swing. Democrats continue to visit the platform but post far less frequently, suggesting a shift from expressive to largely passive participation.

\subsection{Affective Polarization and Posting}

\begin{figure}[H]
\centering
\begin{subfigure}[t]{0.48\textwidth}
    \centering
    \includegraphics[width=\textwidth]{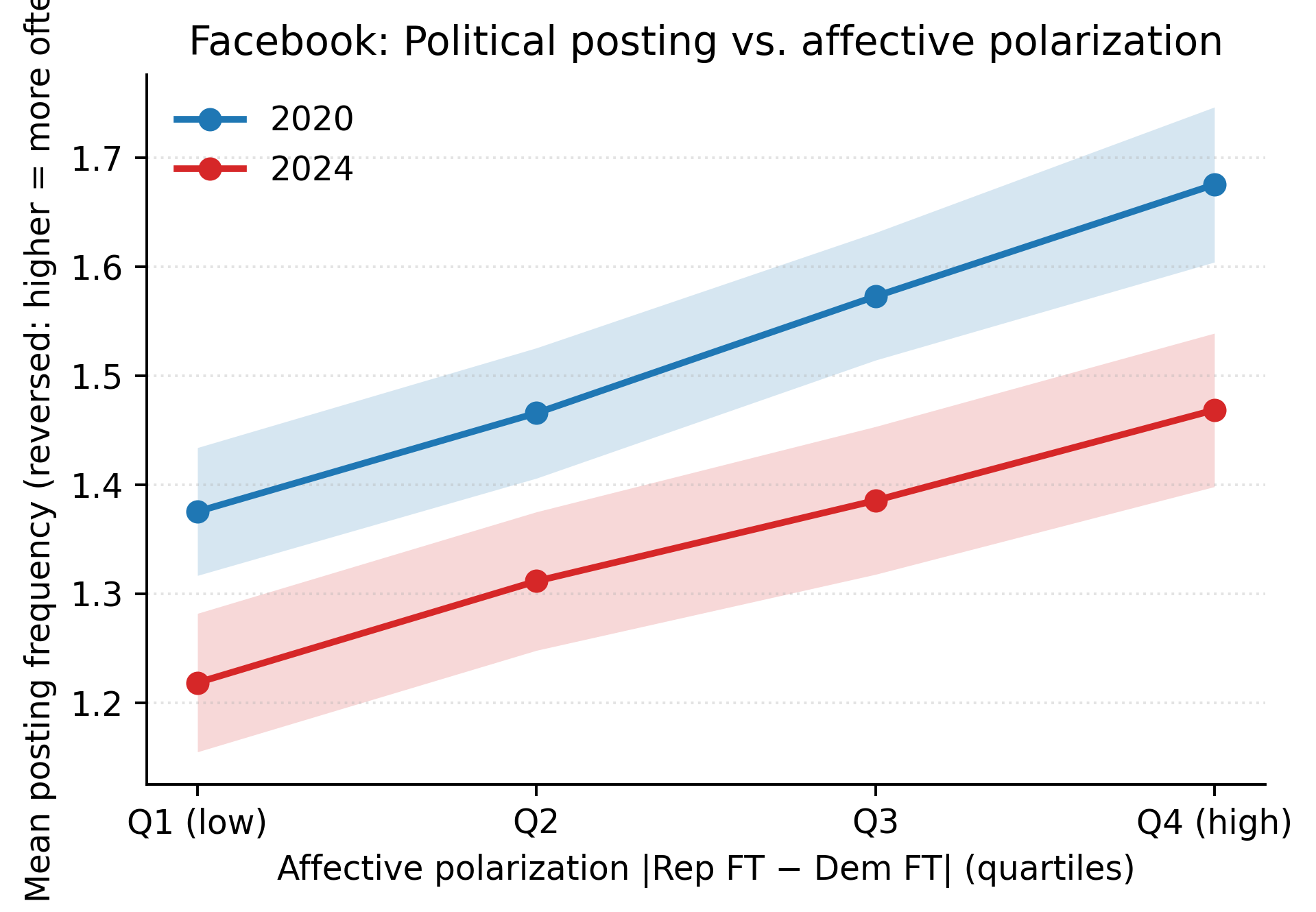}
    \caption{Facebook: political posting frequency by affective polarization, 2020 vs.\ 2024 (ANES, weighted).}
    \label{fig:fb_post_polar}
\end{subfigure}\hfill
\begin{subfigure}[t]{0.48\textwidth}
    \centering
    \includegraphics[width=\textwidth]{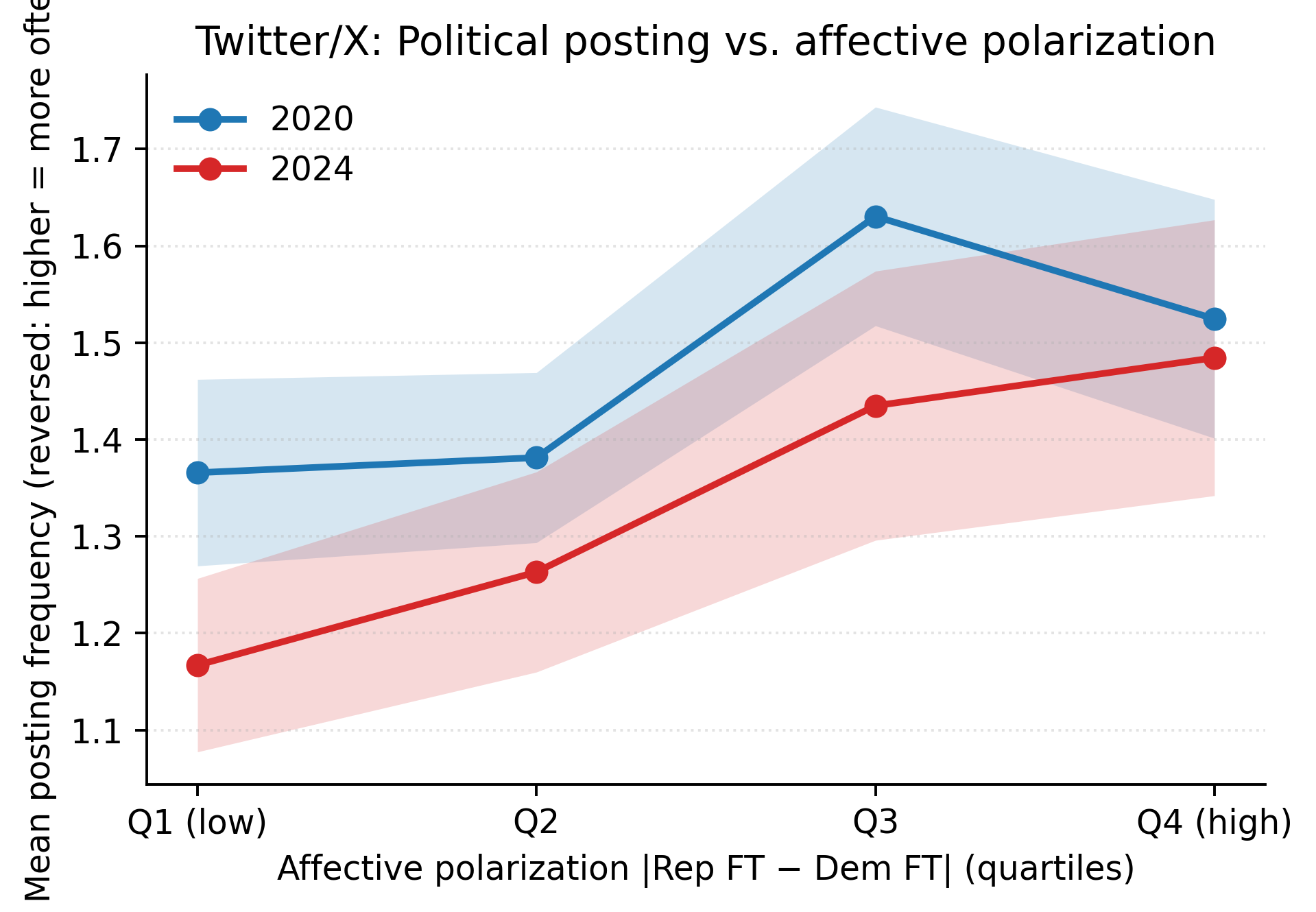}
    \caption{Twitter/X: political posting frequency by affective polarization, 2020 vs.\ 2024 (ANES, weighted).}
    \label{fig:tw_post_polar}
\end{subfigure}
\caption{Posting frequency rises sharply with affective polarization on both platforms. Respondents expressing stronger in-party warmth and out-party hostility post more often about politics. Average posting rates decline in 2024 for all groups, but the positive slope remains—especially on Twitter/X, where polarization is most predictive of activity.}
\label{fig:affpol_post}
\end{figure}

Figures~\ref{fig:fb_post_polar} and~\ref{fig:tw_post_polar} show that political posting remains strongly linked to affective polarization. On both Facebook and Twitter/X, individuals with warmer feelings toward their own party and colder feelings toward the opposing one post more frequently about politics. Between 2020 and 2024, overall posting intensity declines across the board, yet the positive gradient steepens, particularly on Twitter/X. Even as fewer people share political content, those who continue to do so are increasingly drawn from the most polarized segments of the population. As a result, visible political discourse on both platforms becomes dominated by extreme partisans, potentially itself representing a causal mechanism through which social media drive societal polarization \parencite{tornberg2022digital}.

\begin{figure}[H]
\centering
\begin{subfigure}[t]{0.45\textwidth}
    \centering
    \includegraphics[width=\textwidth]{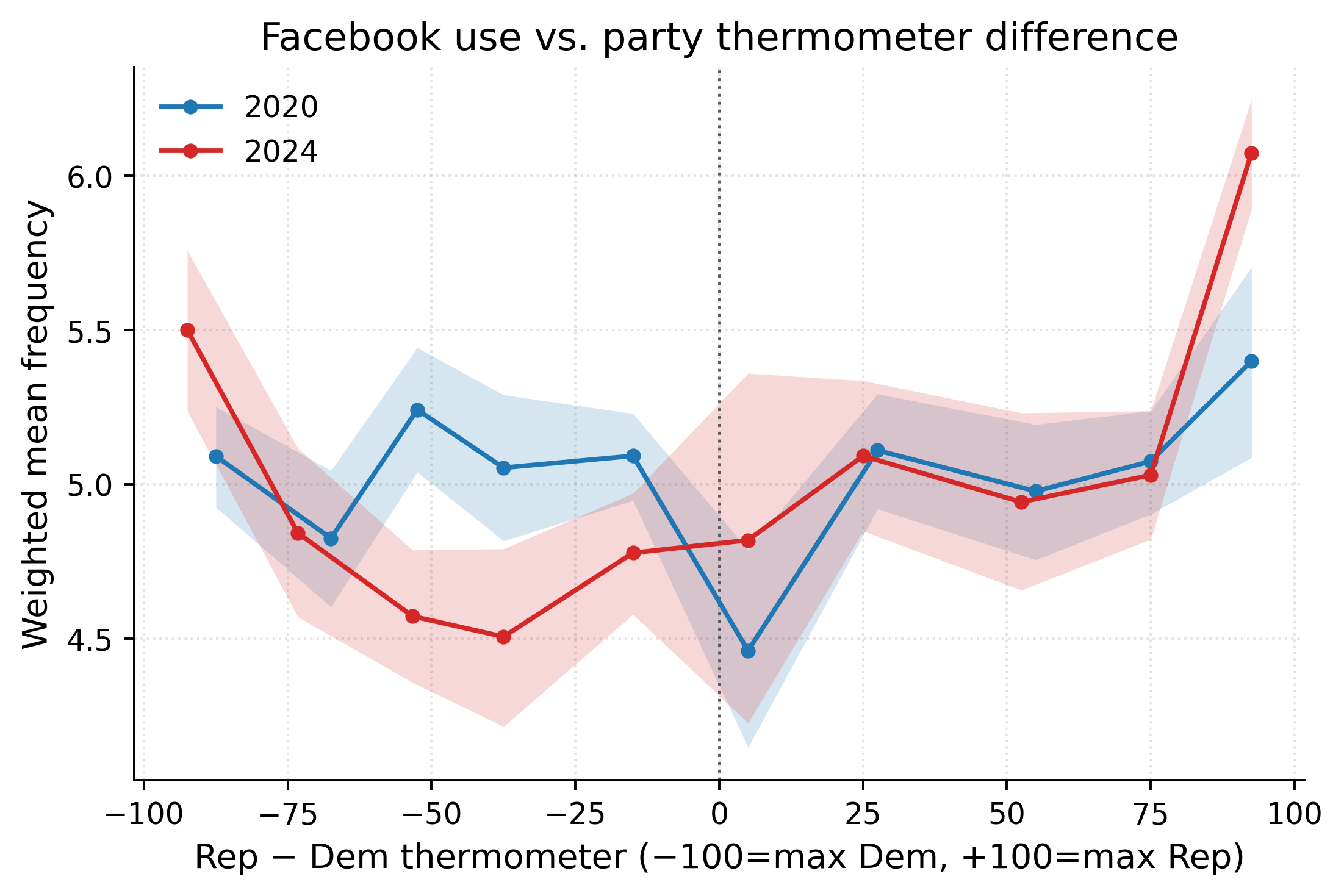}
    \caption{Facebook: frequency of use by Republican affect (party thermometer difference). Strong Republicans remain the heaviest users; the overall level of use declines modestly in 2024 but the upward gradient persists.}
    \label{fig:fb_use}
\end{subfigure}\hfill
\begin{subfigure}[t]{0.45\textwidth}
    \centering
    \includegraphics[width=\textwidth]{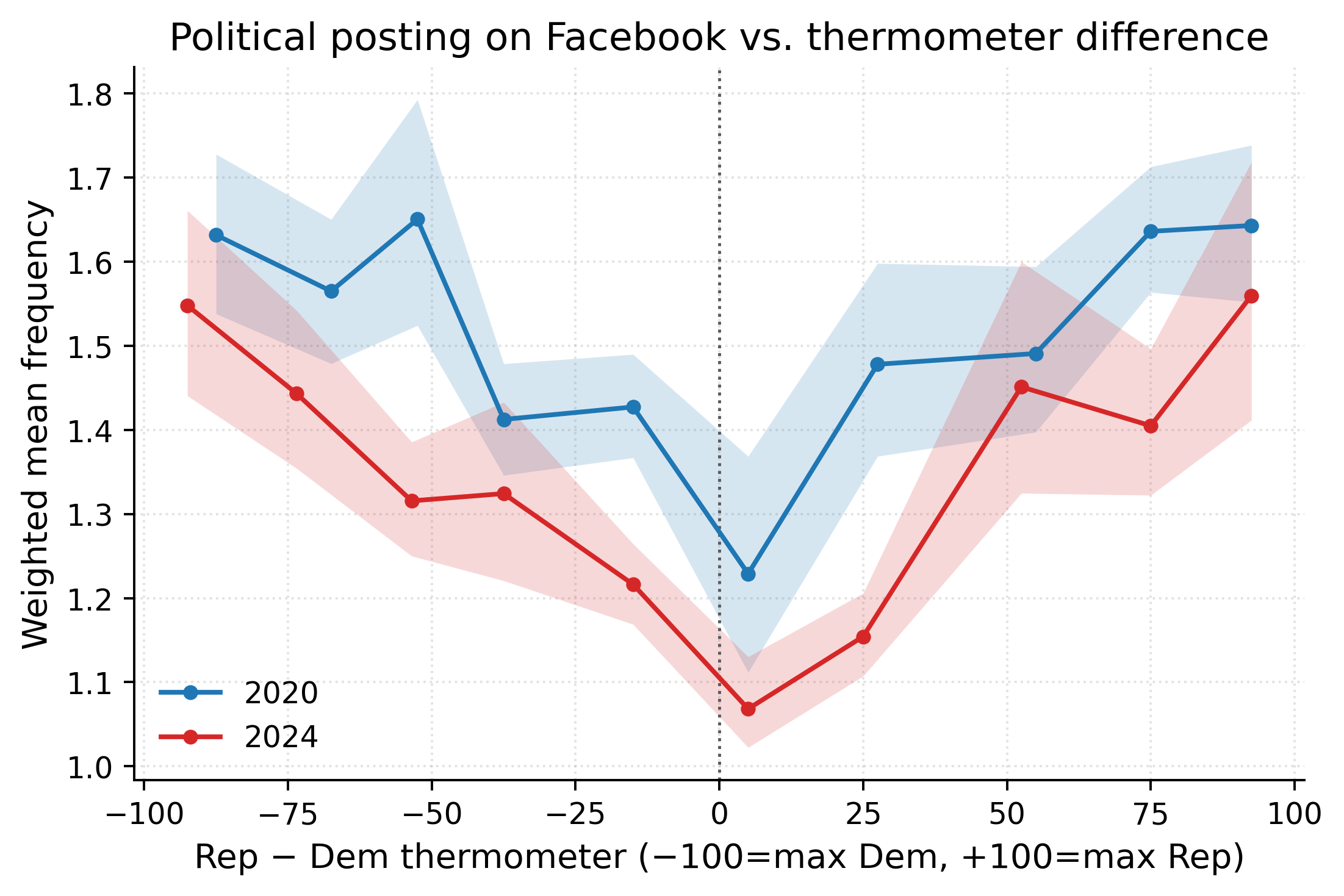}
    \caption{Facebook: posting frequency by Republican affect. Posting declines overall in 2024, yet the tilt toward strong Republicans is stable—those highest in Republican affect continue to post most.}
    \label{fig:fb_post}
\end{subfigure}

\vspace{1em}

\begin{subfigure}[t]{0.45\textwidth}
    \centering
    \includegraphics[width=\textwidth]{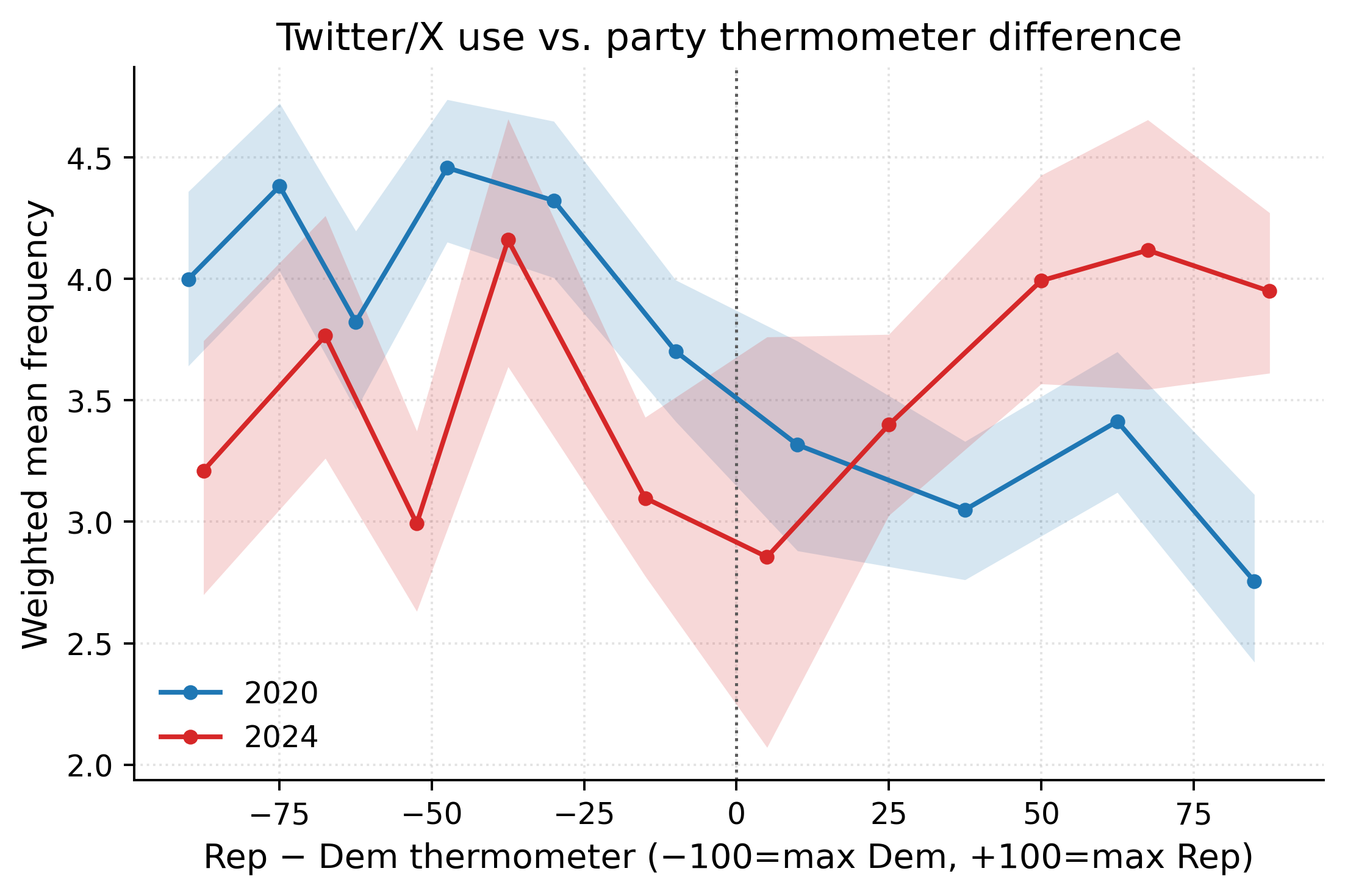}
    \caption{Twitter/X: frequency of use by Republican affect. The platform flips from Democrat-skewed in 2020 to Republican-skewed in 2024, with use rising steadily with Republican affect.}
    \label{fig:tw_use}
\end{subfigure}\hfill
\begin{subfigure}[t]{0.45\textwidth}
    \centering
    \includegraphics[width=\textwidth]{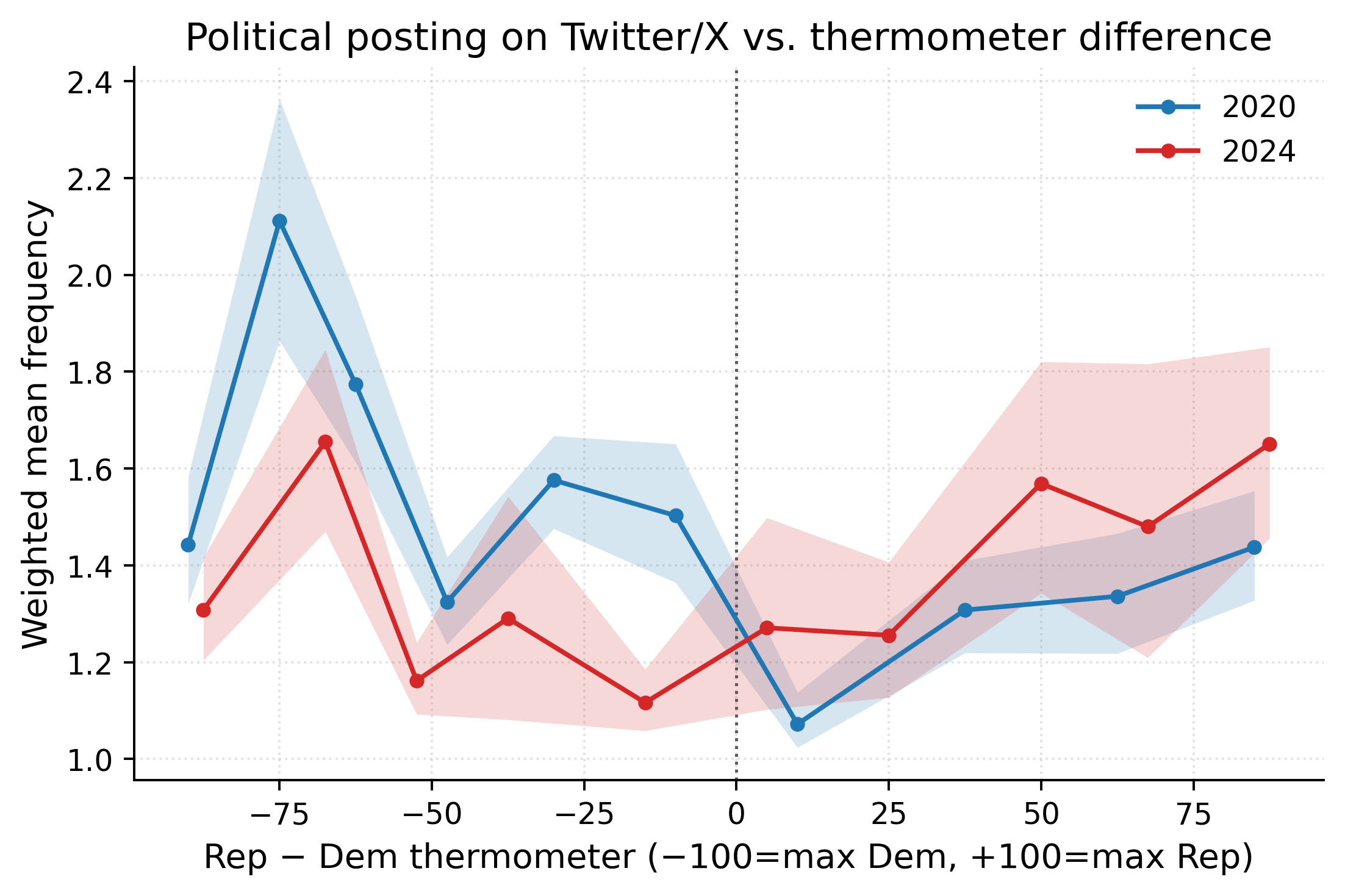}
    \caption{Twitter/X: posting frequency by Republican affect. Posting was highest among Democrats in 2020 but peaks among strong Republicans in 2024; centrists and less polarized users post least.}
    \label{fig:tw_post}
\end{subfigure}

\caption{Association between partisan affect and platform engagement (ANES 2020 vs.\ 2024, weighted). Facebook shows stable Republican-leaning engagement patterns, whereas Twitter/X shifts from a Democratic to a Republican gradient in both use and posting. Polarized users continue to dominate visible political activity even as overall participation declines.}
\label{fig:socialmedia_thermo}
\end{figure}

Figures~\ref{fig:socialmedia_thermo} further illustrate how partisan affect correlates with engagement. On Facebook, both usage and political posting are relatively symmetrical: individuals higher in partisan affect are consistently more active. 

In contrast, Twitter/X exhibits a clear reversal. In 2020, use and posting were concentrated among affectively polarized Democrats, but by 2024 the relationship flips, with engagement now rising with Republican affect. Moderates and respondents expressing ambivalent feelings toward both parties remain the least active on either platform.

Taken together, these results highlight a structural asymmetry in the evolving social media landscape. Overall participation declines, yet the link between affective polarization and posting intensity strengthens -- meaning that the visible public sphere increasingly reflects partisan extremes. The shift of Twitter/X toward Republican-aligned participation amplifies this dynamic: while its user base has become more balanced, those who remain active are disproportionately the most polarized Republican partisans. The result is an online environment with fewer participants but more intense partisan conflict.

\subsection{Demographic Composition by Platform}

\begin{figure}[h!]
\centering

\begin{subfigure}[t]{0.48\textwidth}
    \centering
    \includegraphics[width=\textwidth]{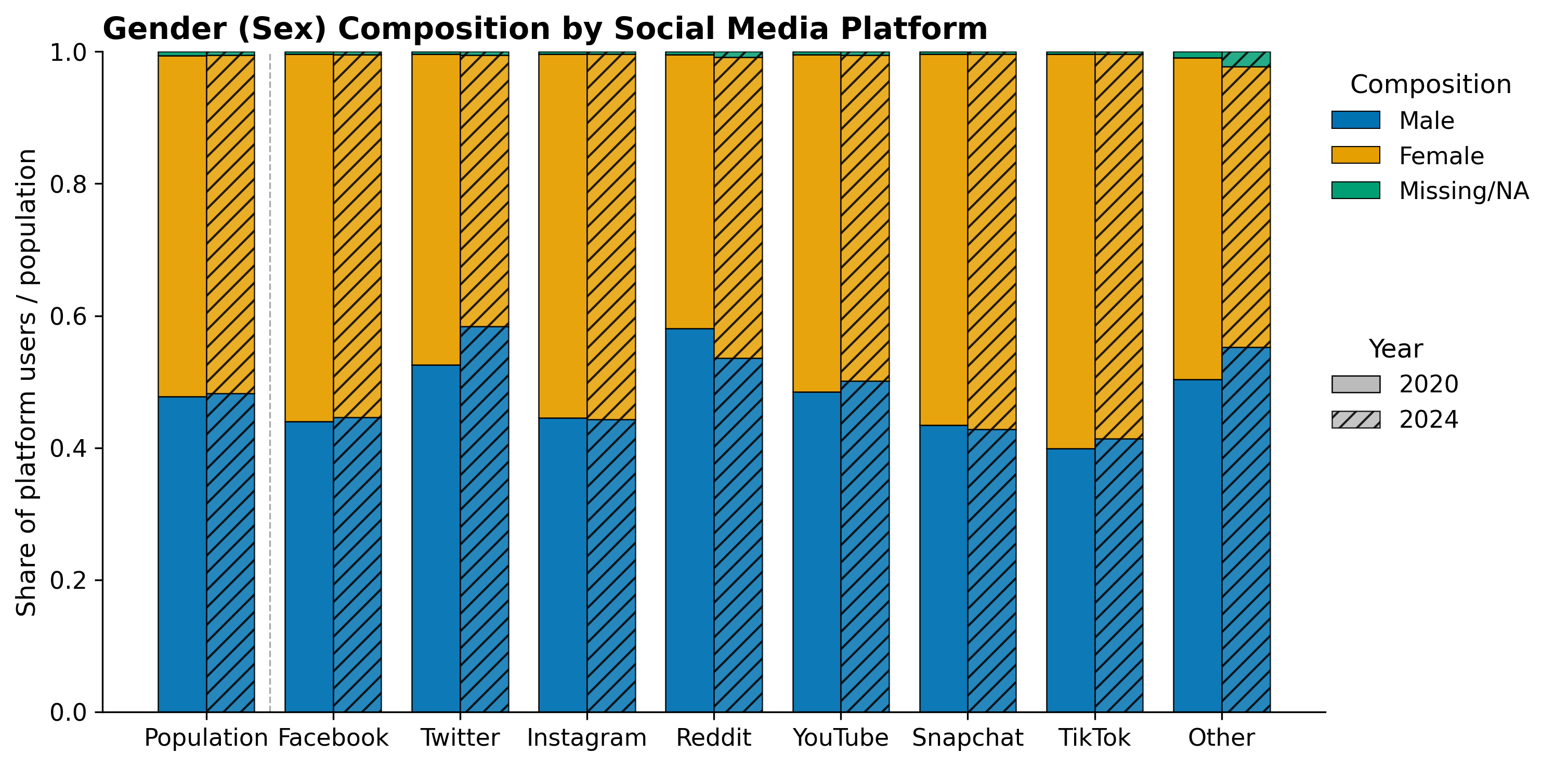}
    \caption{Gender composition by platform. Twitter/X and Reddit remain male-skewed, while Instagram, Snapchat, and TikTok overrepresent women.}
    \label{fig:gender}
\end{subfigure}
\hfill
\begin{subfigure}[t]{0.48\textwidth}
    \centering
    \includegraphics[width=\textwidth]{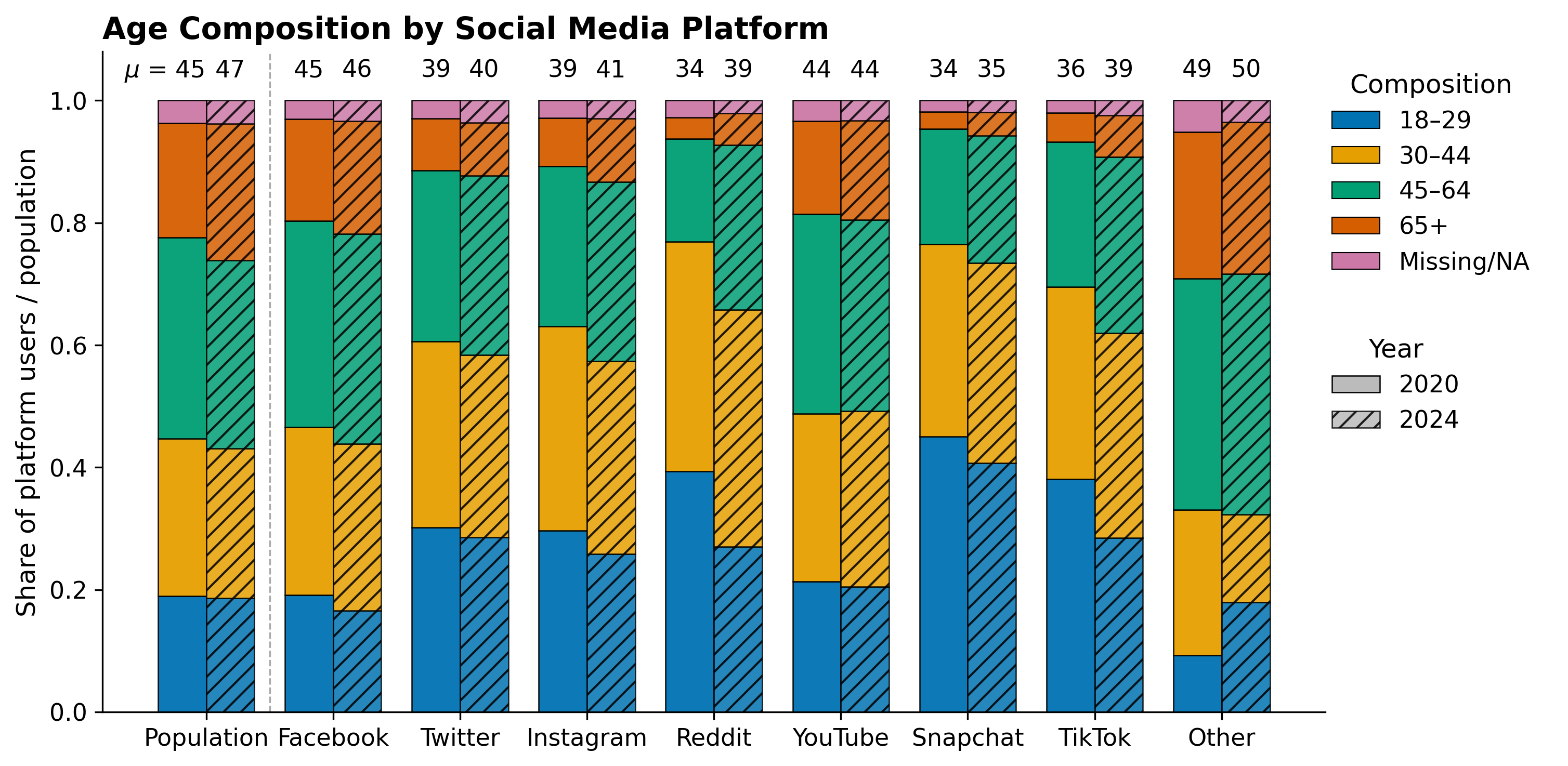}
    \caption{Age composition by platform. Most platforms age modestly, with declines among 18--29 users and growth among middle-aged groups.}
    \label{fig:age}
\end{subfigure}

\vspace{1em}

\begin{subfigure}[t]{0.48\textwidth}
    \centering
    \includegraphics[width=\textwidth]{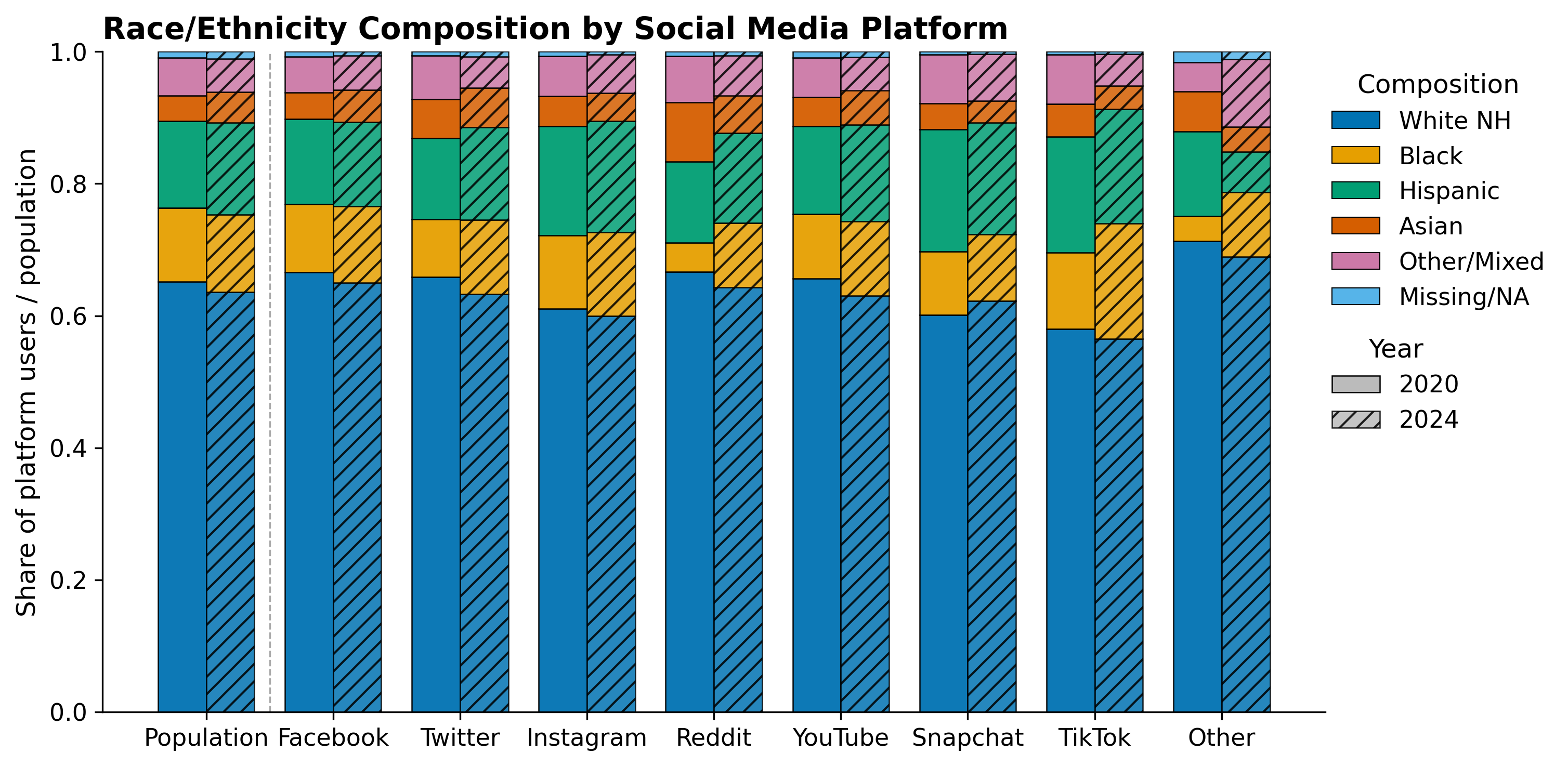}
    \caption{Race and ethnicity composition. Small declines in White users and slight increases among Hispanic and Black users.}
    \label{fig:race}
\end{subfigure}
\hfill
\begin{subfigure}[t]{0.48\textwidth}
    \centering
    \includegraphics[width=\textwidth]{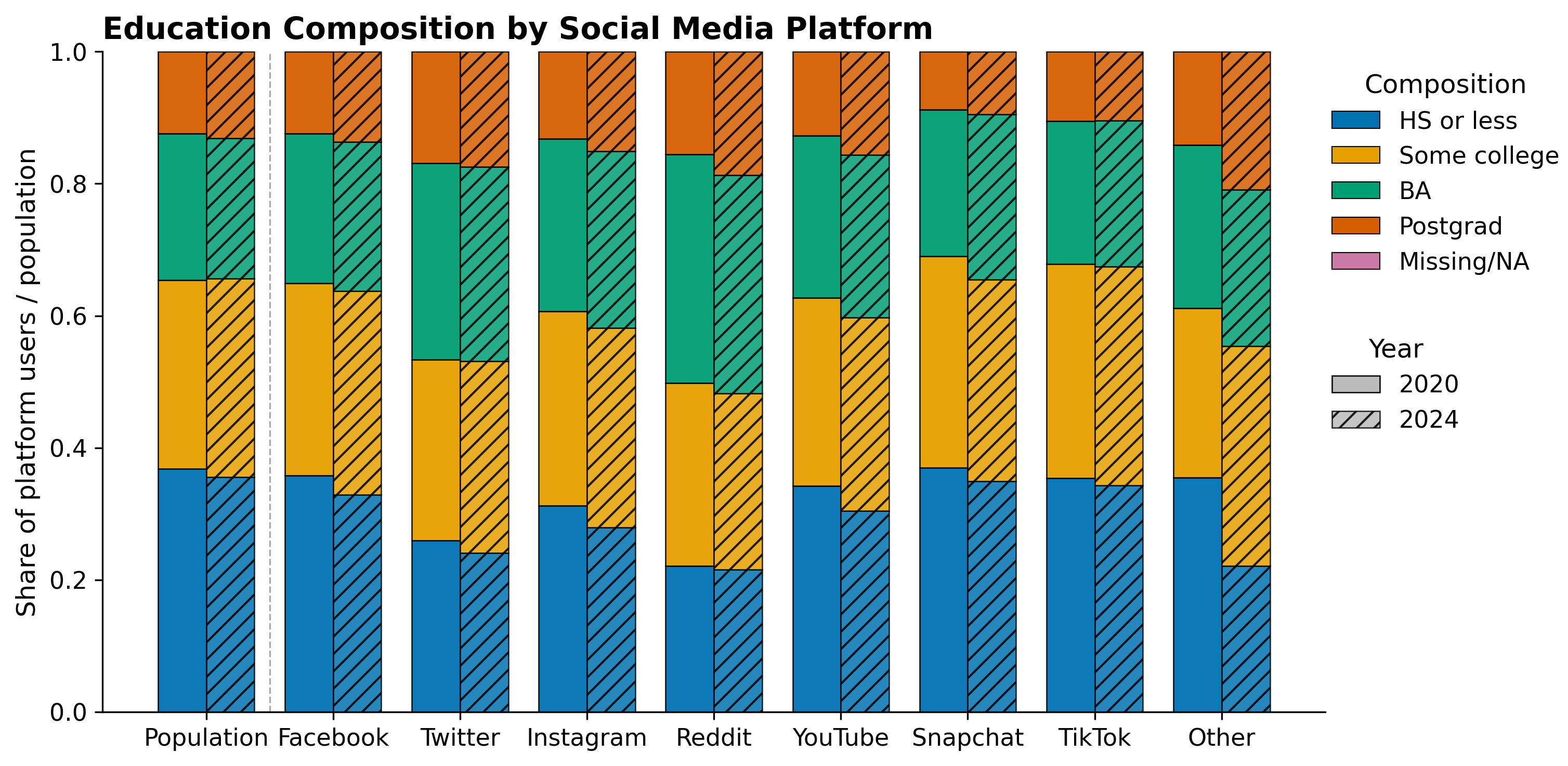}
    \caption{Education composition. Shares of college-educated users rise modestly, especially on Twitter/X and Reddit.}
    \label{fig:educ}
\end{subfigure}

\caption{Demographic composition of major social media platforms, 2020 vs.\ 2024 (ANES, weighted population shares within platform--year; 95\% CIs).  
Across gender, age, race/ethnicity, and education, platform audiences show gradual aging, a mild educational up-tilt, and slight racial diversification, but overall demographic stability.}
\label{fig:demographics_2x2}
\end{figure}

Figure~\ref{fig:demographics_2x2} summarizes how the demographic composition of social media platforms has evolved between 2020 and 2024. Gender patterns are largely stable. Twitter/X has become somewhat more male-skewed, surpassing Reddit as the most male-dominated platform, while Instagram, Snapchat, and TikTok continue to overrepresent women. Facebook and YouTube remain closest to population averages.  

Most platforms have aged modestly, likely reflecting both cohort aging and the selective withdrawal of younger users. The share of 18--29-year-olds declines across TikTok, Instagram, Reddit, and Snapchat, while middle-aged groups expand. Facebook remains the oldest platform, and YouTube continues to mirror the population’s overall age distribution.  

User bases have also diversified slightly in terms of race and ethnicity. White representation has declined marginally on Instagram, YouTube, Twitter/X, and TikTok, accompanied by small increases among Hispanic and Black users. These shifts are modest and vary across platforms, indicating incremental diversification rather than sharp demographic change.  

Educational composition tilts upward across nearly all platforms. Twitter/X and Reddit continue to attract the most highly educated audiences, whereas Facebook and TikTok remain more mixed. Overall, there is a mild but consistent shift toward users with at least a bachelor’s degree, mirroring broader educational trends in internet use.  

Taken together, these trends point to slow but steady demographic evolution: social media audiences are aging slightly, becoming more educated, and diversifying marginally, yet the overall profile of platform users remains remarkably stable over time.

\subsection{Conclusion}
The U.S.\ social media landscape is quietly reshaping itself. Between 2020 and 2024, overall platform use slipped, driven by a rise in the population -- especially the youngest and oldest -- who no longer use social media at all. The old incumbents -- Facebook, YouTube, and Twitter/X -- have lost ground, while TikTok and Reddit have expanded modestly. The users who remain are slightly older, better educated, and more racially diverse than four years ago.

The political balance of social media has shifted just as noticeably. The once-clear Democratic lean of major platforms has declined. Twitter/X, in particular, has seen a radical flip: a space dominated by Democrats in 2020 is now more Republican-aligned, especially among its most active users and posters. Reddit's remains a Democraic stronghold, but its liberal edge has softened. 

Across platforms, overall political posting has declined, yet its link with affective polarization persists. Those expressing the strongest partisan animus continue to post most frequently, meaning that visible political discourse remains dominated by the most polarized voices. This leads to a distorted representation of politics, that itself can function as a driver of societal polarization \parencite{tornberg2022digital,bail2022breaking}. 

Overall, the data depict a social media ecosystem in slow contraction and segmentation. As casual users disengage while polarized partisans remain highly active, the tone of online political life may grow more conflictual even as participation declines. The digital public sphere is becoming smaller, sharper, and louder: fewer participants, but stronger opinions. What remains online is a politics that feels more divided -- not because more people are fighting, but because the fighters are the ones left talking.

\section{Code Availability}
All code and replication materials are openly available at the project’s GitHub repository: \url{https://github.com/cssmodels/anesanalysis}.

\printbibliography

\end{document}